\documentclass[draft]{svjour3}

\usepackage{natbib}
\usepackage{tikz-cd}
\newcommand\bm[1]{\mbox{\boldmath $#1$}}

\def\npk{\bm{k}}
\def\npl{\bm{l}}
\def\npm{\bm{m}}
\def\npmbar{\bm{\overline{m}}}

\def\half{\smfrac{1}{2}}

\def\CKP{Cartan-Karlhede procedure}

\newcommand{\eqref}[1]{(\ref{#1})}
\newcommand{\smfrac}[2]{{\textstyle{#1\over#2}}}

\makeatletter
\newcommand{\ms}{\noalign{\vspace{3\p@ plus2\p@ minus1\p@}}}
\@addtoreset{equation}{section}
\makeatother

\begin{document}

\title{Spacetimes with  continuous linear isotropies II: boosts}

\author{M. A. H. MacCallum}

\institute{M. A. H. MacCallum \at School of Mathematical Sciences,
Queen Mary University of London, Mile End Road, London E1 4NS, UK.
\email{M.A.H.MacCallum@qmul.ac.uk}
}

\maketitle
\begin{abstract}
    Conditions are found which ensure that local boost invariance
    (LBI), invariance under a linear boost isotropy, implies local
    boost symmetry (LBS), i.e.\ the existence of a local group of
    motions such that for every point $P$ in a neighbourhood there is
    a boost leaving $P$ fixed. It is shown that for Petrov type D
    spacetimes this requires LBI of the Riemann tensor and its first
    derivative. That is also true for most conformally flat
    spacetimes, but those with Ricci tensors of Segre type [1(11,1)]
    may require LBI of the first three derivatives of curvature to
    ensure LBS.
\end{abstract}

\section{Introduction}

In the first paper of this series \citep{Mac21a} spacetimes with a
local spatial rotational invariance were (re)-investigated. Here the
corresponding issues for local boost invariance are studied. The issues
arising and the methods to be used are set out in Sections 1 and 2 of
the previous paper. Only those key points required to make this paper
reasonably self-contained will be repeated here.

In studying local rotational symmetry of spacetime (LRS), \cite{Ell67}
introduced three definitions which he showed to be equivalent for dust
spacetimes. The results were later extended to spacetimes with a
perfect fluid and electromagnetic field \citep{SteEll68}. Here Ellis's
definition (A${}_m$) will be studied. It reads:
\begin{quote}
  (A${}_m$) At each point $P$ in an open neighbourhood
$U$ of a point $P_o$, there exists a nondiscrete subgroup $g$ of the
Lorentz group in the tangent space $T_P$ which leaves invariant the
curvature tensor and all its covariant derivatives to the $m$-th
order.
\end{quote}
Implicit in this definition are conditions on the smoothness of the
manifold and the correspondence between the $g$ at separated
points. This paper considers the case where the group $g$ of linear
isotropies contains boosts. This is called local boost invariance
(LBI), `local' meaning that the same $g$ applies throughout $U$.
`Spacetime' here just means a four-dimensional Lorentzian
manifold. The field equations of general relativity will not be used,
but in cases where the Ricci tensor takes the form that would be
implied by specific matter content in general relativity that
interpretation is referred to.

The starting conjecture (based on a claim by \cite{Sik76} now known to
be false in general) is that (A${}_1$) is sufficent to imply Ellis's
definition (C), i.e.
\begin{quote}There exists a local group of motions $G_r$ in an open
neighbourhood $W$ of a point $P_o$ which is multiply transitive on some
$q$[-dimensional] surface through each point $P$ of $W$.
\end{quote}
With LBI this would imply that the spacetime had local
boost symmetry (LBS), i.e.\ that the group $G_r$ contained, for every
$P\in W$, a subgroup of boosts leaving $P$ fixed. A theorem of
\cite{Hal89} implies, under appropriate topological and smoothness
conditions, that $W$ is a region of a manifold in which the same $G_r$
acts globally. 

Only Petrov type D and conformally flat spacetimes have a Weyl tensor
that can satisfy (A${}_0$) (or (A${}_m$) for larger $m$) with a group
$g$ containing boosts.  There are rather more Ricci tensor types that
can satisfy (A${}_0$) for boosts. For (A${}_0$) to apply to the whole
Riemann tensor, the Weyl and Ricci tensors must of course be
appropriately aligned, and there may be a nonzero Ricci scalar.
The tracefree part of the Ricci tensor can be characterized by its
Segre type. The possible invariance groups of the Ricci tensor were
listed by Segre type in Table 5.2 of \cite{SteKraMac03}. Table 1 here
lists those with nontrivial invariance groups $\hat{I}_0$ which include
a boost.
\begin{table}[ht] \tabcolsep0.7ex
  \centering\caption{Nontrivial invariance groups containing boosts,
    by Ricci tensor type.}
  \vspace{2ex}
\label{tab:5.2}
\begin{tabular}{@{}ll@{}} \ms \hline\hline \ms
Invariance group & Segre type of the Ricci tensor \\ \ms\hline \ms
\ms
Boosts (only) & $[11(1,1)]$ \\
Boosts and rotations & $[(11)(1,1)]$ \\
$SO(2,1)$: three-dimensional Lorentz group & $[1(11,1)]$ \\
Full Lorentz group & $[(111,1)]$ \\ \ms
\hline\hline
\end{tabular}
\end{table}
\cite{CahDef68} showed that for Petrov type D spacetimes with boost
(or spatial rotation) invariance and any compatible Ricci tensor,
(A${}_2$) was a sufficient criterion for the spacetime to be LRS or
LBS. Subsequently \cite{GooWai86} gave criteria for the LRS Petrov
type D case in terms of the spin coefficients and curvature expressed
in a Newman-Penrose (NP) null tetrad. These criteria were shown in
\cite{Mac21a} to be equivalent to (A${}_1$). The discrepancy with
Cahen and Defrise's use of (A${}_2$) is shown in the Appendix of
\cite{Mac21a} to be due to a less suitable choice of frame in the
calculations.

Here it is shown in Section \ref{sec:LBS}, by arguments parallel to those
of \cite{GooWai86}, that (A${}_1$) is also sufficient for LBS in
Petrov type D spacetimes. In Section \ref{CFcases}, the corresponding
question for conformally flat spacetimes is studied. In both these
sections the detailed arguments are closely related to those of
\cite{Mac21a} by the asterisk operation of the GHP formalism
\citep{GerHelPen73}. A recent preprint addresses local boost
invariance of higher dimensional manifolds \citep{McNColWyl19}.

As in the previous paper, the \CKP\ for characterizing spacetimes and
testing their equivalence, as outlined in Section 2 of \cite{Mac21a},
is used. It relies on the computation of ``Cartan invariants'', the
components of the Riemann tensor and its covariant derivatives in
canonically chosen frames.  The implementation used here employs the
Newman-Penrose formalism as set out in Chapter 7 of
\cite{SteKraMac03}. The ``Newman-Penrose equations'' (the Ricci
equations) and Bianchi identities [(7.21a)-(7.21r) and (7.32a)-(7.32k)
  in \cite{SteKraMac03}] will be referred to below as (NPa)--(NPr) and
(Ba)--(Bk).

A minimal set of Cartan invariants sufficient for the above procedure,
was defined by \cite{MacAma86}.  It consists of totally symmetrized
spinor derivatives of the Newman-Penrose curvature quantities. Here
the shorthand notation for such spinors, as defined in \cite{Mac21a},
will be used. If $Q^{ABC\ldots}{}_{E'F'\ldots}$ is a relevant
curvature quantity then the notation $Q_{AB'}$ denotes the component
of $Q^{(ABC\ldots)}{}_{(E'F'\ldots)}$ in which $A$ of the $m$ unprimed
indices and $B$ of the $n$ primed indices are contracted with the
basis spinors $\iota$ and $\bar{\iota}$ respectively (and the others
with the basis spinors $o$ and $\bar{o}$). $\chi$ is said to have
valence $(m,\,n)$.  The set defined in \cite{MacAma86} consists of the
totally symmetrized derivatives of $\Psi$, $\Phi$ and $\Lambda$, together
with, at order 1, $\Xi_{DEFW'}={\nabla^C}_{W'}\Psi_{CDEF}$ and at
order $q+2$, the d'Alembertians of quantities at order $q$. For a
totally-symmetrized spinor of valence $(m,\,n)$, only components with
$2(A+B)=m+n$ are LBI.

\section{Petrov type D spacetimes with local boost invariance}
\label{sec:LBS}
In the calculations, the boost invariance is assumed to act in the
$(\npk,\,\npl)$ plane of a Newman-Penrose tetrad adapted to the Petrov
type D Weyl tensor and to leave the Riemann tensor and its first
derivative unchanged\footnote{The results however are independent of
  the choice of frame: they would simply be more difficult to check in
  a randomly chosen frame.}. $\Lambda$ and $\Psi_2 (\neq 0)$ are boost
invariant, so from (A${}_1$) $D\Lambda = \Delta \Lambda =
D\Psi_2=\Delta\Psi_2=0$. In $\Phi_{AB'}$ only the components
$\Phi_{11'}$ and $\Phi_{02'}$ can be nonzero. Boost invariance of the
Cartan invariants $\nabla{\Psi_{AB'}}$ requires that
 $$\nabla{\Psi_{01'}} =  \nabla{\Psi_{11'}} = \nabla{\Psi_{20'}} =
\nabla{\Psi_{31'}} = \nabla{\Psi_{40'}} =\nabla{\Psi_{41'}} =0 $$
and thus
\begin{equation}\label{C1dagvalues}
  \kappa = \sigma = \rho = \mu = \lambda = \nu =0.
\end{equation}
These spacetimes are members of Kundt's class.

Therefore, in Petrov type D, invariance of the Riemann tensor and its
first derivatives under a boost implies that there is a Newman-Penrose
tetrad (a canonical one for Petrov type D, fixed up to a spatial
rotation and boost) in which the following criteria hold.
\begin{eqnarray}
  {\rm (C1}{}^\dag): &&  \kappa=\sigma=\rho=\mu=\lambda=\nu=0.\\
  {\rm (C2}{}^\dag): &&  \Phi_{01}=\Phi_{12}=0.\\
  {\rm (C3}{}^\dag): &&  \Delta \Lambda= D\Lambda = 0.
\end{eqnarray}
{}[The (Cn${}^\dag$) notation is adopted to emphasize the parallel with
the conditions (C1)--(C3) of \cite{GooWai86}.]
The theorem analogous to Theorem 2.1 of Goode and Wainwright is:
\begin{theorem} \label{LBSThm1}
A space-time $($assumed conformally curved$)$ is LBS if
and only if there exists a null tetrad $(\npk,\,
  \npl,\,\npm,\,\npmbar)$ in which
{\rm (C1${}^\dag)-($C3${}^\dag$)} hold.
\end{theorem}

To show that conversely (C1${}^\dag$)--(C3${}^\dag$) imply that the
Riemann tensor and its first derivatives are boost invariant, one
first inserts (C1${}^\dag$) into (NPa, b, j, k, m and n), and, using
(C2${}^\dag$), obtains that
$$\Phi_{00'}=\Phi_{22'} =0, \qquad \Psi_0=\Psi_1=\Psi_3=\Psi_4=0,$$ so
that the Riemann tensor is Petrov type D or conformally flat and is
boost invariant. (Be) and (Bf) give $D\Psi_2=\Delta\Psi_2=0$. From
above, (C1${}^\dag$) and (C3${}^\dag$) then guarantee that $\nabla
\Lambda$ and $\nabla \Psi_{AB'}$ are boost invariant.

Using (C1${}^\dag$)--(C3${}^\dag$) in (Bi) and (Bk) gives $D\Phi_{11'}
=\Delta\Phi_{11'}=0$, so the gradient of $\Phi_{11'}$ is boost
invariant.  The components of $\nabla\Phi$ that must vanish for LRSI
are given in full as \eqref{DPhi01B}--\eqref{DPhi23B} in Section
\ref{CFcases}, and the relations between these components of
$\nabla\Phi$ and the Bianchi identities are discussed there. Given
(C1), they all vanish, due to (Bb) and (Bc), so $\nabla\Phi$ is boost
invariant.

To complete the check of the equivalence of the conditions
(C1${}^\dag$)-(C3${}^\dag$) with the assumption that the Riemann
tensor and its first derivatives are boost invariant, one has to show
that once the remaining frame freedom, a spatial rotation, has been
fixed, so that $\Phi_{02'}$ is an invariant,
$D\Phi_{02'}=\Delta\Phi_{02'} =0$, which follows if $\varepsilon$ and
$\gamma$ are real.  (The boost invariance of $\Xi_{AB'}$ is readily
checked.)

These and other restrictions on the spin coefficients analogous to
those in Section 3 of \cite{Mac21a} are now sought, following
analogous steps in \cite{GooWai86}.  They will enable the LBI of
higher derivatives of the curvature to be checked.  From (Bh) and (Bj)
one finds
$$ \delta(-\Psi_2-\Phi_{11'}+\Lambda) = -3\tau\Psi_2 + 2\bar{\pi}\Phi_{11}
+\pi\Phi_{02'}.$$
Applying the $[\bar{\delta},\, D]$ commutator to
$(-\Psi_2-\Phi_{11'}+\Lambda)$ and using (NPc) [which tells us that
  $D\tau=(\varepsilon-\bar{\varepsilon})\tau$], (Bb) and $\Psi_2\neq
0$ one obtains $D\pi= -(\varepsilon-\bar{\varepsilon})\pi$.

Similarly, from (Bg) and the conjugate of (Bj) one obtains
$$ \bar{\delta}(-\Psi_2-\Phi_{11'}+\Lambda) = 3\pi\Psi_2 -
2\bar{\tau}\Phi_{11} - \tau\Phi_{20'}.$$ Applying the
commutator
$[\bar{\delta},\,\Delta]$ to $(-\Psi_2-\Phi_{11'}+\Lambda)$, using (NPi)
[which tells us $\Delta \pi = -(\gamma-\bar{\gamma})\pi$], (Bc) and
$\Psi_2 \neq 0$ one obtains $\Delta\tau = (\gamma-\bar{\gamma})\tau$.

Thus from (NPc), (NPi), and the arguments above:
\begin{eqnarray}\label{dpitau1}
  D\pi= -(\varepsilon-\bar{\varepsilon})\pi, &\quad&
 \quad \Delta\tau =
 (\gamma-\bar{\gamma})\tau \\
 \label{dpitau2}
 D\tau=(\varepsilon-\bar{\varepsilon})\tau, &\quad& \Delta \pi =
-(\gamma-\bar{\gamma})\pi.
\end{eqnarray}
Note that to arrive at \eqref{dpitau1} in conformally flat cases
one would need to derive it by a different argument.

It will now be shown that a position-dependent rotation (the
remaining frame freedom) can always be used to achieve
\begin{equation}\label{aplusb2}
  \varepsilon=\bar{\varepsilon} \quad {\rm and} \quad
  \gamma=\bar{\gamma},
\end{equation}
and hence that
\begin{equation} \label{DDeltas}
  D\Phi_{02'} = \Delta \Phi_{02'}=D\pi = \Delta\pi=D\tau=\Delta\tau=0.
\end{equation}

For non-zero $\tau$ and $\pi$, if either $\pi$ or $\tau$ is real
\eqref{DDeltas} follows immediately from
\eqref{dpitau1}--\eqref{dpitau2}.  If $\pi$ and $\tau$ are both
non-zero $D(\ln(\tau/\pi))=2(\varepsilon-\bar{\varepsilon})$ and
$\Delta(\ln(\tau/\pi))= 2(\gamma-\bar{\gamma})$. If then $\tau/\pi$ is
real, these imply \eqref{aplusb2}. Under a rotation through an angle
$\theta$,
$(\varepsilon-\bar{\varepsilon})^\star=(\varepsilon-\bar{\varepsilon})+iD\theta$
and $(\gamma-\bar{\gamma})^\star=(\gamma-\bar{\gamma})+i\Delta\theta.$
Hence if $\tau/\pi$ is not real, $\theta=\half {\rm Im}(\ln
(\pi/\tau))$ achieves \eqref{aplusb2}.
If just one of $\pi$ or $\tau$ is nonzero, its argument gives a
suitable $\theta$. If $\pi=0=\tau$ one needs to show that the equations
\begin{equation}\label{thetaeqs}
  iD\theta = -(\varepsilon -\bar{\varepsilon})\quad {\rm and} \quad
  i\Delta\theta = -(\gamma -\bar{\gamma}),
\end{equation}
are compatible. This is done by applying the  $[\Delta,\,D]$
commutator to $\theta$. Note that (NPh) implies that
$\Psi_2$ is real in this case; the imaginary part of (NPf) then
shows that the two equations for $\theta$ are indeed compatible.

Finally, (NPe) and the complex conjugate of (NPd), and (NPr)
and the complex conjugate of (NPo), are used to obtain
\begin{equation}\label{aminusb}
  D(\alpha - \bar{\beta}) = 0 = \Delta(\alpha-\bar{\beta}).
\end{equation}
Inserting these results into the higher derivatives, using CLASSI,
shows that the second, third and fourth derivatives of the Riemann
tensor are also boost invariant. This will also be true in conformally
flat spacetimes if \eqref{C1dagvalues}, \eqref{aminusb} and
\eqref{DDeltas} hold, but in the case studied in Section
\ref{SO21case} (A${}_3$) is required to establish \eqref{DDeltas}.

One can show, as in the LRS case, that in Petrov type D one cannot
have $s=1$ and $t_2=1$, $t_1=t_0=0$ which would require (A${}_4$) to
be checked. To eliminate the
possibility, the calculations follow a similar logic to those in
\cite{Mac21a}. Necessarily $\Psi_2 \neq 0$, and $\Phi_{11'}$ and
$\Phi_{02'}$ would be constant. That $\nabla\Psi_{AB'}$ is constant
implies $\tau$ and $\pi$ are constant. If at least one of them is
nonzero, then (NPg) and/or (NPp) imply that $\bar{\alpha}-\beta$ is
constant (possibly zero). Direct calculation (using CLASSI) then shows
$t_2=0$ i.e.\ all terms in the second derivatives are also constant,
so the \CKP\ terminates.  If both $\pi$ and $\tau$ are zero, (NPg)
implies $\Phi_{02'}=0$, and then inspection (using CLASSI) shows that
all first derivatives of the Riemann tensor, and hence all higher
derivatives, are zero, and the \CKP\ terminates at step 1.

Thus for LBS Petrov type D spacetimes it is sufficient to check
(A${}_3$) and the \CKP\ must terminate at the third step or
earlier. This proves the following.

\begin{theorem} \label{LBSDthm2}
  If a spacetime of Petrov type D is such that the Riemann tensor and
  its first derivative are invariant under a local boost invariance,
  then the spacetime is locally boost symmetric and admits a local
  isometry group $G_r~(r\geq 3)$.
\end{theorem}

The converse of Theorem \ref{LBSDthm2} is obvious, and by the
equivalence shown above this proves Theorem \ref{LBSThm1}. Note that
as in Section 3 of \cite{Mac21a} the invariance of the derivatives of
the Ricci tensor has not been used to derive the results, only
checked, and $\Psi_2 \neq 0$ was used only in deriving (C1${}^\dag$)
and \eqref{dpitau1}.  The LBS conclusion depends only on
\eqref{C1dagvalues}, \eqref{aminusb} and \eqref{DDeltas}.

In the following section it is shown that Theorem \ref{LBSDthm2} is
still true with `Petrov type D' replaced by 'conformally flat', unless
the Ricci tensor is of Segre type [1(11,1)] when some cases require
LBI of the curvature and its first three derivatives to ensure
LBS.

\section{ Conformally flat spacetimes with local boost invariance}
\label{CFcases}

The conformally flat cases to be considered are those Ricci tensor
types appearing in Table 1.  (A${}_m$) is assumed to hold with a group
$g$ which contains a boost. By the same argument as in \cite{Mac21a},
Ricci tensors of Segre type [(111,1)] are easily disposed of: the
spacetimes are of constant curvature, the subgroup $g$ in (A${}_m$) is
the trivial one comprising the whole Lorentz group, $s=6$,
$t_p=0=t_0$, and there is a group $G_{10}$ transitive on the whole
spacetime. The \CKP\ terminates at the first step and (A${}_0$)
suffices because it will imply (A${}_1$).

In the rest of this section the actual (A${}_m$) required for local
LBS in conformally flat spacetimes with the various Ricci tensors
which admit a boost invariance, but are of less symmetry than Segre type
[(111,1)], are studied. 

\subsection{The first derivatives and Bianchi identities}
\label{CF1stDerivs}

The first step is to impose LBI on the first derivatives of $\Phi$ and
$\Lambda$. Then one can try to derive (C1${}^\dag$), which were
obtained in Petrov type D cases from invariance of $\nabla\Psi$, and
look for an appropriate variant of the rest of the arguments in Section
\ref{sec:LBS}.

With LBI of the curvature only $\Lambda$, $\Phi_{11'}$ and
$\Phi_{02'}=\overline{\Phi_{20'}}$ can be nonzero components in a
suitable canonically chosen frame, and since $\Phi_{11'}$ and
$\Lambda$ are invariant under the remaining allowed changes of frame,
(A${}_1$) requires that $\Delta\Phi_{11'}=D\Phi_{11'}=\Delta \Lambda =
D\Lambda = 0$. (Note that the invariance of curvature is being assumed
here, rather than $\Phi_{00'}=\Psi_{22'}=0$ being deduced from other
assumptions as in Section \ref{sec:LBS}.) The terms in $\nabla\Phi$
that must vanish if (A${}_1$) holds are
\begin{eqnarray}
3\nabla\Phi_{01'}&=&
2(2\kappa\Phi_{11'}+\bar{\kappa}\Phi_{02'}), \label{DPhi01B}\\
3\nabla\Phi_{02'}&=&
D\Phi_{02'}+2(\bar{\rho}-\varepsilon+\bar{\varepsilon})\Phi_{02'}
+4\sigma\Phi_{11'},\label{DPhi02B}\\
9\nabla\Phi_{11'}/2&=& 2(\rho+\bar{\rho})\Phi_{11'}+ \sigma\Phi_{20'}+\bar{\sigma}\Phi_{02'},\label{DPhi11B}\\
  3\nabla\Phi_{13'}&=&\Delta\Phi_{02'} + 2(\bar{\gamma}-\gamma-\mu)\Phi_{02'}-4\bar{\lambda}\Phi_{11'},\label{DPhi13B}\\
  9\nabla\Phi_{22'}/2&=&-2(\mu+\bar{\mu})\Phi_{11'}-\lambda\Phi_{02}
-\bar{\lambda}\Phi_{20'}.\label{DPhi22B}\\
3\nabla\Phi_{23'}&=&
-2(\nu\Phi_{02'}+2\bar{\nu}\Phi_{11'}).\label{DPhi23B}
\end{eqnarray}
In addition $\nabla\Phi_{00'} = \nabla\Phi_{33'}\equiv 0$ here.  Of
the above equations, \eqref{DPhi01B} is equivalent to (Ba),
\eqref{DPhi11B} to (Bi) or the real part of (Be), \eqref{DPhi22B} to
(Bk) or the real part of (Bf), and \eqref{DPhi23B} to the conjugate of
(Bd). The Bianchi identities (Be) and (Bf) give
\begin{equation}\label{BeAndBf}
  \bar{\sigma}\Phi_{02'}+2\rho\Phi_{11'}=0,\quad {\rm and} \quad
  \bar{\lambda}\Phi_{20'}+2\mu\Phi_{11'}=0.
\end{equation}
In the general case the only information from boost invariance
of $\nabla\Phi$ additional to that in the Bianchi identities comes
from \eqref{DPhi02B} and (Bb), and \eqref{DPhi13B} and the conjugate
of (Bc), which give respectively.
\begin{equation}
  \label{DPhiRemB}
  \bar{\rho}\Phi_{02'}+2\sigma\Phi_{11'}=0,\quad {\rm and} \quad
  \bar{\mu}\Phi_{20'}+2\lambda\Phi_{11'}=0.
\end{equation}
If \eqref{DPhiRemB} is satisfied the first derivatives of the
curvature are boost invariant. (In the special case of Segre type
[1(11,1)], where $D\Phi_{02'}=\Delta\Phi_{02'}=0$, \eqref{DPhiRemB} is
equivalent to (Bb) and (Bc), and contains no additional
information. This case is considered in Section \ref{SO21case} below.)

The Bianchi identities (Bg) and (Bh) give
\begin{equation}\label{pitau}
  2(\bar{\pi}+\tau)\Phi_{11'}+(\pi +\bar{\tau})\Phi_{02'}=0.
\end{equation}
The remaining information in the Bianchi identities is
\begin{eqnarray}\label{CFBg}
  Bg: & \delta\Phi_{20'} -2\bar{\delta}\Lambda=&
  (2\bar{\alpha}-2\beta-\bar{\pi})\Phi_{20'}-2\pi\Phi_{11'},\\
  \label{CFBj}
  Bj:& -\delta\Phi_{11'}-\bar{\delta}\Phi_{02'}+3\delta \Lambda = &
  (-2\alpha+2\bar{\beta}+\pi-\bar{\tau})\Phi_{02'}\\
  && \phantom{-2\alpha}+2(\bar{\pi}-\tau)\Phi_{11'}. \nonumber
\end{eqnarray}

\subsection{Ricci tensors of Segre type [11(1,1)]}
\label{ssec:CFG1}

If such spacetimes obey (A${}_\infty$) the \CKP\ must terminate after
at most 3 steps since $s=1$ at every step and at most two steps are
needed to increase $t_p$ to the maximum of 2. So (A${}_3$) would
suffice. As shown next, only (A${}_1$) is actually required. All the
resulting spacetimes admit a $G_r,~r\geq 3$ acting on submanifolds of
dimension at least 2.

The ratio $\Phi_{02'}/\Phi_{11'}=2c$ is fixed under the remaining
frame freedom but may be position-dependent. (Using the remaining
rotational freedom one could make $\Phi_{02'}$ real and positive, and
$c$ real, but this freedom may be needed as in Section \ref{sec:LBS}
to obtain a frame in which \eqref{aplusb2} holds, using
\eqref{dpitau1}--\eqref{dpitau2}.) One must have $|c| \neq 1$ or the
Segre type will be [1(11,1)]. Relations of the form
$2\Phi_{11'}Q+\Phi_{02'}\bar{Q}=0$ imply $Q=0$, since otherwise
$|Q/\bar{Q}|=1\neq |c|=|\Phi_{02'}/\Phi_{11'}|$ (If $|c|=1$ at some
isolated points, continuity implies $Q=0$ there too.) Hence from (Ba)
and (Bd), $\kappa=\nu=0$, and from \eqref{DPhiRemB} and
\eqref{BeAndBf} $\rho=\mu=\sigma=\lambda=0$. So (C1${}^\dag$)
holds. In addition, \eqref{pitau} implies $\bar{\pi}+\tau=0$.

\eqref{dpitau2} holds (being just (NPc) and (NPi)) and \eqref{dpitau1}
then follows simply from it, using $\bar{\pi}+\tau=0$. One can then
obtain \eqref{aplusb2} and \eqref{DDeltas}, and complete the proof
that (A${}_1$) is sufficient to imply LBS in this case as in Section
\ref{sec:LBS}. One may note that since
$D\Phi_{02'}=\Delta\Phi_{02'}=0$, $c$ is constant in the timelike
two-planes determined by the boost.

\subsection{Ricci tensors of Segre type [(11)(1,1)]}

This Segre type is that of the Ricci tensor of a non-null
electromagnetic field.  Here $s_0=2$ and in a canonical frame
$\Phi_{02'}=0$.  Assuming $\Phi_{11'}\neq 0$, (Ba)--(Bd) give
$$\kappa=\sigma=\lambda=\nu=0.$$ If $s=2$ one must have $t_1=0$ and so
the \CKP\ terminates at step 1 since neither $s$ nor $t$ has changed
and as in Section 3 of \cite{Mac21a} this gives the Bertotti-Robinson type
solutions with a $G_6$ transitive on the whole spacetime, and
(A${}_1$) suffices.

For $s=1$ with $\hat{I}_1$ consisting of the boosts,
$D\Lambda=D\Phi_{11'}=\Delta \Lambda = \Delta\Phi_{11'}=0$ and from
(Be) and (Bf) $\rho=\mu=0$ so the conditions of Theorem \ref{LBSThm1}
hold.  As in the previous subsection, one has $\tau+\bar{\pi}=0$, so
the proof that (A${}_1$) with a one-dimensional group of boosts
implies LBS proceeds as in section \ref{sec:LBS}.

\subsection{Ricci tensors of Segre type [1(11,1)]}
\label{SO21case}

Here the invariance group is SO(2,1), generated by null rotations
about $\npk$ and about $\npl$ and a boost in the $(\npk,\,\npl)$ plane.  The
SO(2,1) group acts in a hyperplane and leaves invariant one direction
in the $(\npm,\,\npmbar)$ plane. The Ricci tensor represents a
tachyonic fluid, and a canonical form for it which is manifestly null
rotation invariant (as in \cite{Mac20}) about each of the null
directions has only $\Phi_{11'}$ and $\Phi_{02'}$ non-zero with
$2|\Phi_{11'}|=|\Phi_{02'}|$. (This is a specialization of the form
for Segre type [11(1,1)], treated above.) Using the remaining freedom
of spatial rotation in the $(\npm,\,\npmbar)$ plane one can set
$\Phi_{02'} =2\Phi_{11'}$: the parameters of both null rotations are
then pure imaginary and the vector orthogonal to the hyperplane in
which the SO(2,1) acts is in the direction $\npm+\npmbar$.

The boost rescales the parameters of the null rotations. One might
therefore have invariance under a two-dimensional subgroup of SO(2,1)
generated by the boost and one of the null rotations (see entry R6 in
Table 6.1 of \cite{Hal04}). Among the quantities defined by
\cite{MacAma86}, only $\nabla^k\Phi_{AB'}$ and
$\nabla^k\Lambda_{AB'}$, both of which are Hermitian, and
d'Alembertians thereof, have to be considered. Boost invariance
implies that of the components of $\nabla^k\Phi_{AB'}$, only those
with $A+B=2+k$ can be nonzero. If then $\nabla^k\Phi_{AB'}$ is
invariant under one of the null rotations, then from the Hermitian
symmetry it will also be invariant under the other. Thus
$\nabla^k\Phi_{AB'}$ will be SO(2,1) invariant. The same applies to
$\nabla^k\Lambda_{AB'}$ and the d'Alembertians of spinors of lower
derivative order.

So $s=2$ is impossible and either $s=3$ or $s=1$. If $s=3$ there can
be at most one independent function of position (as there is just one
spacelike direction fixed under $g$). The spacetime admits a $G_6$ (or
in special cases a $G_7$, cf.\ \cite{RebTei92}) acting on timelike
hypersurfaces of constant curvature. The metrics include analogues of
the FLRW metrics for perfect fluids.

The remaining case is where $s=1$ and $g$ in (A${}_m$) is just the boost
invariance\footnote{It might be convenient to carry out the necessary
  calculations using the 3+1 orthonormal tetrad formalism based on a
  spacelike congruence introduced by \cite{Har82}.}. From (Ba) and
(Bd)--(Bf) one has
\begin{equation}\label{CFtacBian}
\kappa+\bar{\kappa}=\nu+\bar{\nu}=\bar{\mu}+\lambda =\bar{\rho}+\sigma=0,
\end{equation}
and then (Bb) and (Bc) give \eqref{aplusb2}.  In
this case the Bianchi identities and
$D\Phi_{11'}=\Delta\Phi_{11'}=D\Lambda=\Delta \Lambda = 0$ ensure (A${}_1$),
\eqref{DPhiRemB} being equivalent to \eqref{BeAndBf} in this case.
From \eqref{pitau} one has
\begin{equation}\label{ptident}
  (\pi+\bar{\pi})+ (\tau+\bar{\tau})=0.
\end{equation}

The real parts of the left sides of \eqref{CFBg} and \eqref{CFBj} are
proportional to the real part of
$\delta(\Phi_{11'}-\Lambda)$.
For the right sides to be
compatible one must have
\begin{equation}
  \label{abpident} 0=(\alpha+\bar{\alpha})-(\beta+\bar{\beta})+(\pi+\bar{\pi}).
\end{equation}

Eliminating between the imaginary parts of \eqref{CFBg} and \eqref{CFBj} yields
\begin{eqnarray}
  \label{delR}(\delta-\bar{\delta})\Lambda&=&[(\bar{\alpha}-\beta)-(\alpha-\bar{\beta})]\Phi_{11'},  \\
  \label{delPhi}
  (\delta-\bar{\delta})\Phi_{11'}&=&[(\bar{\alpha}-\beta)-(\alpha-\bar{\beta})]\Phi_{11'}.  
\end{eqnarray}
The perhaps surprising equality of the right sides does not imply that
$\delta(\Phi_{11'}-\Lambda)=0$.

The real parts of (NPc) and (NPi) give $D (\tau+\bar{\tau})=0=
\Delta(\pi+\bar{\pi})$, so
\begin{equation}\label{dpt}
  D (\tau+\bar{\tau})=\Delta(\pi+\bar{\pi}) = \Delta
  (\tau+\bar{\tau})=D(\pi+\bar{\pi})=0.
\end{equation}
From the imaginary part of (NPg) minus the conjugate of (NPh), using
\eqref{CFtacBian} and \eqref{abpident},
\begin{equation}\label{delpi}
  \delta(\pi+\bar{\pi})=  \bar{\delta}(\pi+\bar{\pi}).
\end{equation}

From the conjugate of (NPd) together with (NPe), and from the
conjugate of (NPr) with (NPo), one finds using \eqref{abpident} that
\begin{equation}
  D(\bar{\alpha}-\beta)= \Delta(\bar{\alpha}-\beta)= 0,
\end{equation}
which is consistent with \eqref{abpident} and \eqref{dpt}.

From (NPb) and the complex conjugate of (NPa), and from the conjugate
of (NPj) and (NPn), one obtains
\begin{equation}\label{kident}
  \kappa(2\bar{\alpha}-2\beta+(\pi + \bar{\pi}))=0=
  \nu(2\bar{\alpha}-2\beta+(\pi + \bar{\pi})).
\end{equation}
Since $\kappa$ and $\nu$ are pure imaginary, the imaginary parts of
these equations follow from \eqref{abpident} while the real parts give
$\kappa\Im_{\alpha\beta}=\nu\Im_{\alpha\beta}=0$, where
$\Im_{\alpha\beta} \equiv
(\alpha-\bar{\beta})-(\bar{\alpha}-\beta)$. (From the relations
between null and orthonormal tetrads obtainable using \cite{GooWai86}
and \cite{Har82} one finds that $\Im_{\alpha\beta}$ plays the same
r\^ole here as $\dot{u}_1$ did in the conformally flat perfect fluids
studied in \cite{Mac21a}.) So far only (A${}_1$) has been used.

Calculating $\nabla^2\Phi$ shows that the $00'$, $01'$, $34'$ and
$44'$ components are identically zero. Most of the other components
with $A+B\neq 4$ vanish on simplification with the help of
\eqref{delPhi}, \eqref{ptident}, \eqref{abpident}, and \eqref{kident},
which all follow from (A${}_1$). The remaining components which must
vanish for LBI ($\nabla^2\Phi_{03}$ and $\nabla^2\Phi_{14'}$) yield
$\Im_{\alpha\beta}\sigma= \Im_{\alpha\beta}\mu =0$. For
the spacetime to obey (A${}_2$) thus requires $\Im_{\alpha\beta}$
multiplying each of $\kappa$, $\sigma$, $\mu$ and $\nu$ to be zero.

To satisfy (A${}_2$) there are now two possibilities: either (a)
$\Im_{\alpha\beta}=(\delta-\bar{\delta})\Phi_{11'}
=(\delta-\bar{\delta})\Lambda=0$, showing that $\Phi_{11'}$ and
$\Lambda$ are constant in the hyperplanes defined by the SO(2,1)
action, or (b) $\Im_{\alpha\beta}\neq 0$ and (C1${}^\dag$) holds.

In case (a), we find that $\nabla^k\Phi$ and $\nabla^k\Lambda$ are
invariant under a null rotation about $\npk$ for $k=1\dots3$ and
therefore $s=3$ by the earlier argument that $s_i$ cannot be 2.  We
thus again have at least a $G_6$ if $\Im_{\alpha\beta}=0$. Note that
(A${}_2$) was checked but does not give extra conditions in this case.

In case (b), one now has the conditions (C1${}^\dag$) and
\eqref{aplusb2}. (NPc) and (NPi) give $D\tau=\Delta\pi=0$. To complete
\eqref{DDeltas}, one has to show that $D\pi=\Delta \tau=0$.  Then the
curvature derivatives up to the fourth will have boost invariance, as
in Section \ref{sec:LBS}. (Note that since
$(\delta-\bar{\delta})\Phi_{11'}\neq 0$, $t_0=0$ need not be
considered here.)  One finds $\nabla^3\Phi_{22'}=
-6\Phi_{11'}\Im_{\alpha\beta}D(\pi+\tau)/25$ and $\nabla^3\Phi_{33'}$
gives the same with $\Delta$ replacing $D$. Thus to obtain
\eqref{DDeltas} one needs (A${}_3$) in this case.

\section*{Acknowledgements}
I am grateful to Jan {\AA}man for his work in developing the software
CLASSI and to Filipe Mena, who was kind enough to check and correct some of the
calculations.

\section{Conclusion}

The work in Sections \ref{sec:LBS} and \ref{CFcases} gives
the following analogue of Theorem 3 of \cite{Mac21a}.

\begin{theorem} In spacetimes with a Ricci tensor of Segre type
    $[1(11,1)]$ whose distinguished spacelike eigenvector is not geodesic, local
  boost invariance of the curvature and its derivatives
  up to the third holds if and only if the spacetime is locally
  boost symmetric. In all other cases, local boost
  invariance of the curvature and its first derivatives holds if and
  only if the spacetime is locally boost symmetric.
\end{theorem}


\end{document}